\documentclass[amsmath,amssymb,preprint,showpacs]{revtex4}
\usepackage{graphicx}
\usepackage{bm}

\newcommand{\be}{\begin{eqnarray}}
\newcommand{\ee}{\end{eqnarray}}

\begin{document}
\title{Cartoon Computation: Quantum-like computing without quantum mechanics}
\author{Diederik Aerts $^1$ and Marek Czachor $^{1,2}$}
\affiliation{
$^1$ Centrum Leo Apostel (CLEA) and Foundations of the Exact Sciences (FUND)\\
Vrije Universiteit Brussel, 1050 Brussels, Belgium\\
$^2$ Katedra Fizyki Teoretycznej i Informatyki Kwantowej\\
Politechnika Gda\'nska, 80-952 Gda\'nsk, Poland}

\begin{abstract}
We present a computational framework based on geometric structures.
No quantum mechanics is involved, and yet the algorithms perform tasks analogous to quantum computation. Tensor products and entangled states are not needed --- they are replaced by sets of basic shapes. To test the formalism we solve in geometric terms the
Deutsch-Jozsa problem, historically the first example that
demonstrated the potential power of quantum computation. Each step of the algorithm has a clear geometric interpetation and allows for a cartoon representation.
\end{abstract}

\maketitle

\section{Introduction} 

Thinking of quantum computation one typically has in mind a quantum computer --- a device based on and limited by the laws of the microworld. But the same laws that allow for quantum computation state that the noise occuring in actual systems may make the computation more or less unrealistic. The two recent US and UE strategic reports \cite{rep} show how the level of practical difficulties varies from implementation to implementation. The goal of our paper is to show that perhaps one should also look for non-microworld implementations of quantum computation. More precisely, we present a framework for quantum-like algorithms that does not refer to quantum mechanics, and involves only geometric structures algebraized by means of geometric algebras (GA). 
To prove that the new framework indeed works we solve in a GA
way the celebrated Deutsch--Jozsa (DJ) problem \cite{DJ}.

There are some trivial ways of including GA in quantum computation
\cite{foot}, but this is not what we want to do. The GA algorithm we
present below is not just a simple translation of the quantum one.
As opposed to quantum computation the basic operation is not the
tensor product but the geometric (Clifford) product (the map that
forms an oriented square from two vectors, a cube from a vector and
a square, a square from a vector and a cube, and so on, as shown on
the figures). In quantum computation we are bound to use unitary
operations (quantum dynamics is unitary) and projectors
(measurements are represented by projections). Binary numbers are
built by  tensoring qubits with one another. All these operations
are higly counterintuitive. 

In GA computation the operations are different and there is nothing
counterintuitive about them. After a necessary amount of exercise
the operations can be visualized without great difficulty. Binary
numbers are coded directly in terms of basic geometric shapes, with
no tensor product or quantum entanglement. Parallel processing is
performed on `bags of shapes'. In GA computation one can really {\it
see\/} the solution. 

\section{Elements of GA} 

Let us now recall those basic facts about GA
that are important for our purposes \cite{GA1,GA2,GA3,H1,H3,Pavsic}.
One begins with an $n$-dimensional Euklidean space $V_n$ whose
orthonormal basis is $\{e_1,\dots,e_n\}$. The associative geometric
product $ab$ of two vectors $a=\sum_{k=1}^na_k e_k$,
$b=\sum_{k=1}^nb_k e_k$, is defined by linearity from the Clifford
algebra \be e_ke_l+e_le_k=2\delta_{kl}\bm 1 \ee of the basis. Here
$\bm 1$ is the neutral element of the GA: $a\bm 1=\bm 1 a=a$.
Directed subspaces are then associated with the set of {\it
blades\/} defined as geometric products of different basis vectors
supplemented by the identity $\bm 1$, corresponding to  the basic
oriented scalar (analogous to a charged point). The blades include
vectors (oriented line segments), bivectors (oriented plane
segments), trivectors (oriented volume segments), and so on. A
general element $A$ of GA, called a multivector, is a linear
combination of the blades \be A=A_0\bm 1+\sum_{k} A_{k}
e_{k}+\sum_{k<l} A_{kl} e_{k}e_{l}+\dots+A_{1\dots n} e_1\dots e_n,
\nonumber \ee where the coefficients are real.

Figures~1--2 explicitly illustrate the geometry behind multivectors and their
geometric products in a plane. Geometrically the basic blades and their negatives in 2D are: $\bm 1=\circ$, $e_1=\rightarrow$, $e_2=\uparrow$, $e_{12}=\Box$, $-\bm 1=\bullet$, $-e_1=\leftarrow$, $e_2=\downarrow$, $-e_{12}=\blacksquare$. 
Here $\circ$ and $\bullet$ denote the two oppositely `charged' points. 

Note that the dimension of the space of shapes
associated with the plane is four. Multivectors are `bags of shapes'
and the high dimensionality is similar to the one known from
configuration spaces in mechanics. As one does not have problems with
imagining a $3N$-dimensional space representing
configurations of $N$ particles, there is no difficulty
with visualizing the 4-dimensional space representing the `bags' in
Figure~2.

The first element that seems new and is beyond the standard presentation of GA is
the binary parametrization of blades and the role it plays for the geometric product. Denote: 
$\bm 1=e_{0\dots 0}$,
$e_1=e_{10\dots 0}$, $e_2=e_{010\dots 0}$, $\dots$,
$e_{125}=e_{110010\dots 0}$, $\dots$, $e_{12\dots n-1,n}=e_{11\dots
11}$. The notation shows that there is a one-to-one relation between
an $n$-bit number and an element of GA based on $V_n$. Geometric
product in the binary parametrization reads \cite{ACDM,AC06} 
\be
e_{A_1\dots A_n}e_{B_1\dots B_n} &=&
(-1)^{\sum_{k<l}B_kA_l}e_{(A_1\dots A_n)\oplus(B_1\dots B_n)},
\label{GAr} 
\ee 
where $(A_1\dots A_n)\oplus(B_1\dots B_n)$ means
componentwise addition mod 2, i.e. the $n$-dimensional XOR.  The
geometric product may be thus regarded as a  projective representation of XOR. This observation is the deperture point
for our GA computational framework. 

\section{DJ problem in GA framework} 

Assume $f: \{0,1\}^n\to\{0,1\}$
is a constant or balanced function. Consider an $(n+1)$-dimensional
Euclidean space $V_{n+1}$ with orthonormal basis $\{e_1,\dots
e_{n+1}\}$ and its associated GA. Let $E_{n+1}$ denote the sum of
all the blades, 
\be 
E_{n+1} &=& \sum_{A_1\dots A_{n+1}}e_{A_1\dots A_{n+1}}. 
\ee 
Employing (\ref{GAr}) we find, for $e_{n}=e_{0\dots 0 10}$,
\be 
E_{n+1} e_{n}
&=& \sum_{A_1\dots A_{n+1}}(-1)^{A_{n+1}} e_{A_1\dots A_{n+1}}. 
\ee
This step is analogous to the first step of the DJ quantum algorithm
\cite{DJ}. Indeed, let $U_{n+1}$ be the tensor product of $n+1$
Hadamard gates. Then one begins with 
\be 
U_{n+1}|0\dots 01\rangle
&=& \frac{1}{\sqrt{2^{n+1}}} \sum_{A_1\dots A_{n+1}}
(-1)^{A_{n+1}}|A_1\dots A_{n+1}\rangle.\nonumber 
\ee 
Note the
difference in location of 1 in $e_{0\dots 0 10}$ and $|0\dots 01\rangle$. Now assume there exists an oracle $E_f$ that performs
\be 
E_fe_{A_1\dots A_{n}A_{n+1}} 
&=& e_{A_1\dots A_{n}A_{n+1}}e_{0\dots 0f(A_1\dots A_{n})} 
\ee 
The action of the oracle reduces either to the multiplication by the
$(n+1)$th basis vector $e_{n+1}=e_{0\dots 01}$, if $f(A_1\dots
A_{n})=1$, or to the trivial multiplication by $\bm 1=e_{0\dots 0}$
in the opposite case. 
Accordingly 
\be 
E_fE_{n+1} e_{n} 
&=&
\sum_{A_1\dots A_{n}} (-1)^{f(A_1\dots A_{n})} ( e_{A_1\dots A_{n}0}
- e_{A_1\dots A_{n}1} ).\nonumber 
\ee 
This step is analogous to the
oracle action in the quantum algorithm, where 
\be 
U_fU_{n+1}|0\dots
01\rangle &=& \frac{1}{\sqrt{2^{n+1}}} \sum_{A_1\dots A_{n}}
(-1)^{f(A_1\dots A_{n})}
\Big( |A_1\dots A_{n}0\rangle - |A_1\dots A_{n}1\rangle
\Big) \nonumber 
\ee 
The final step is performed by means of 
\be
F_{n+1} &=& \sum_{A_1\dots A_{n}}e_{A_1\dots A_{n}0}^{\dag}, 
\ee
essentially the reverse \cite{reverse} of $E_n$: 
\be
F_{n+1}E_fE_{n+1} e_{n} &=& \sum_{A_1\dots A_{n}=0}^1
(-1)^{f(A_1\dots A_{n})} e_{0\dots 0} +\dots,\nonumber 
\ee 
where the dots stand for the combination of all the blades different from 
$\bm 1=e_{0\dots 0}$. If $\Pi$ projects on  $e_{0\dots 0}$, then
\be 
{\Pi}F_{n+1}E_fE_{n+1}e_{n} 
&=& \left\{
\begin{array}{cl}
(-1)^{f(0\dots 0)} 2^{n}\bm 1 & {\rm for\,constant}\,f\\
0 & {\rm for\, balanced}\,f
\end{array}
\right. 
\nonumber 
\ee 
Looking at the $e_{0\dots 0}$ component we
conclude that $f$ is constant if the component is nonzero, and
balanced if the component is zero. We have achieved the same goal as
the quantum algorithm.

\section{Cartoon algorithms} 

Cartoon versions of the 2-bit GA algorithm are shown on Figures 3--5. All the three figures involve the same first step, but the oracles are different. The projection $\Pi$ means that we select from the final bag the dots. If the dots are black the function is constant with $f(0)=1$; white dots mean constant function, $f(0)=0$ (the oracle is then trivial), and no dots means zero, i.e. a balanced function. Figure~6 shows the algorithm for the 3-bit problem and constant $f(00)=0$. We do not show the oracle since in this case, analogously to Figure~4, the oracle acts trivially. 
For three bits there are 8 blades: 1 scalar, 3 edges $e_k$, 3 walls $e_{kl}$, and 1 cube 
$e_{123}$. The walls are white on one side, and black on the other. $e_{123}$ is white, and its negative is black. We recommend the readers to translate the cartoon into a GA expression.

\section{Advantages and limitations} 

The basic factor that limits practical applicability of cartoon computation is how to physically implement $E_f$, $E_{n+1}$, $F_{n+1}$. The same problem occurs in standard quantum computation but one hopes that any finte unitary transformation can be physically realized by means of a quantum system, at least in principle. However, assuming that black boxes that perform $E_f$, $E_{n+1}$, $F_{n+1}$ in single runs exist, we obtain the same complexity of the algorithm as the quantum one. The main advantage is that we no longer have to look for such implementations in the microscopic domain. We only need geometry, and not necessarily that of an Euclidean space. Another advantage is that the notions of superposition and entanglement have here a clear geometric meaning (bags of blades) and no tensor products are needed. The coefficients occuring in our `entangled states' do not have a probabilistic meaning, but can be both positive and negative, and thus lead to interference effects. The latter property, in addition to parallelism, is the main feature making our algorithms similar to the quantum ones.

\section{Discussion}

The algorithms could also be represented in a
matrix way with $n$-bit numbers given by Cartan's representation of an appropriate Clifford algebra \cite{BT}. Examples of such calculations can be found in \cite{ACDM,AC06}. The exercise is instructive but can be conceptually 
misleading. As often stressed in the GA literature, the matrix representations introduce  redundant elements that obscure the actual geometric meaning of GA operations. In particular, Cartan's representation is constructed by means of tensor products of Pauli matrices, a fact that may make the impression we are using quantum algorithms
in notational disguise, perhaps extended by unphysical operations, which
is not the case. 

A lot of inspiration for our own work came from certain `quantum-like' constructions known in semantic analysis and artificial intelligence (AI) \cite{AC}. For example, 
the idea of replacing tensor product representations \cite{Smolensky} by their `compressions'
based on alternative multiplications occurs in convolution based distributed representations of cognitive structures
\cite{Plate}. `Bags of shapes' are
analogous to `bags of words' from latent semantic analysis (LSA)
\cite{LSA}.  The links to AI and, more generally to the studies of human intelligence become especially intriguing if one thinks of geometric product as a way of {\it decoding\/} relations between geometric objects. Indeed, consider the following problem: If $
\rightarrow\blacksquare=
\Box\rightarrow=\leftarrow\Box=\downarrow 
$
then
$\blacksquare\leftarrow=?$ The GA solution is
$\blacksquare\leftarrow=\uparrow\rightarrow\leftarrow=\uparrow\bullet=-\uparrow=\downarrow
=\rightarrow\blacksquare
$. 
The computation is based on the observation that $\blacksquare$ should be identified with $\uparrow\rightarrow$, $\Box$ with $\rightarrow\uparrow$, $\blacksquare$ with 
$-\Box$, and $\rightarrow\leftarrow$ with $-$.
Similarity to certain IQ tests is striking.
Of particular relevance to our approach are the binary spatter codes
(BSC) \cite{Kanerva,ACDM}, a method of coding and processing distributed information, based on appropriately defined superpositions of XORed binary strings.

An interesting exercise is to try to imagine the three steps of the algorithm from Figure~6 as three different levels of geometric relations between the three `bags of shapes' representing $F_3$, $E_3$ and $e_2$. After some training one indeed starts to {\it understand\/} and {\it see\/} why the multiplication $F_3E_3e_2$ looks like the rightmost bag. In this way we have approached the intriguing problem of understanding via visualization, and the role played in this context by quantum geometry \cite{Penrose2}.

To conclude, we seem to be dealing with a new research field on the borderland between 
`quantum structures' and cognitive science. The counterintuitive elements typical of quantum computation are here absent. The basic structure is geometric and thus very general. How to build a `parallel geometric processor' based on a physical process is a separate issue, but one is no longer confined to quantum systems. In particular, realizations based on classical physics cannot be excluded. 
Finally, implications for `the missing science of consciousness', artificial intelligence, and various representations of cognitive structures can be far reaching and should be further studied. 

\acknowledgments

We acknowledge the support  of the Flemish
Fund for Scientific Research (FWO Project No. G.0452.04), and the
Polish Ministry of Scientific Research and Information Technology
(solicited) project PZB-MIN 008/P03/2003.

\begin{figure}
\includegraphics[width=8 cm]{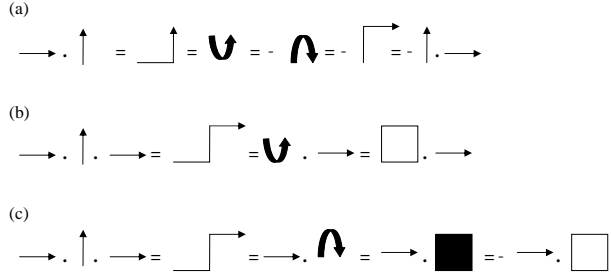}
\caption{Geometric product is noncommutative. (a) Geometry behind
$e_1e_2=e_{12}=-e_2e_1$. (b) Associativity implies
$e_1e_2e_1=e_{12}e_1$. (c) The same as (b) but performed in a
different order. Self-consistency implies that
$e_1e_2e_1=e_{1}(-e_{12})=-e_1 e_{12}$.}
\end{figure}
\begin{figure}
\includegraphics[width=7 cm]{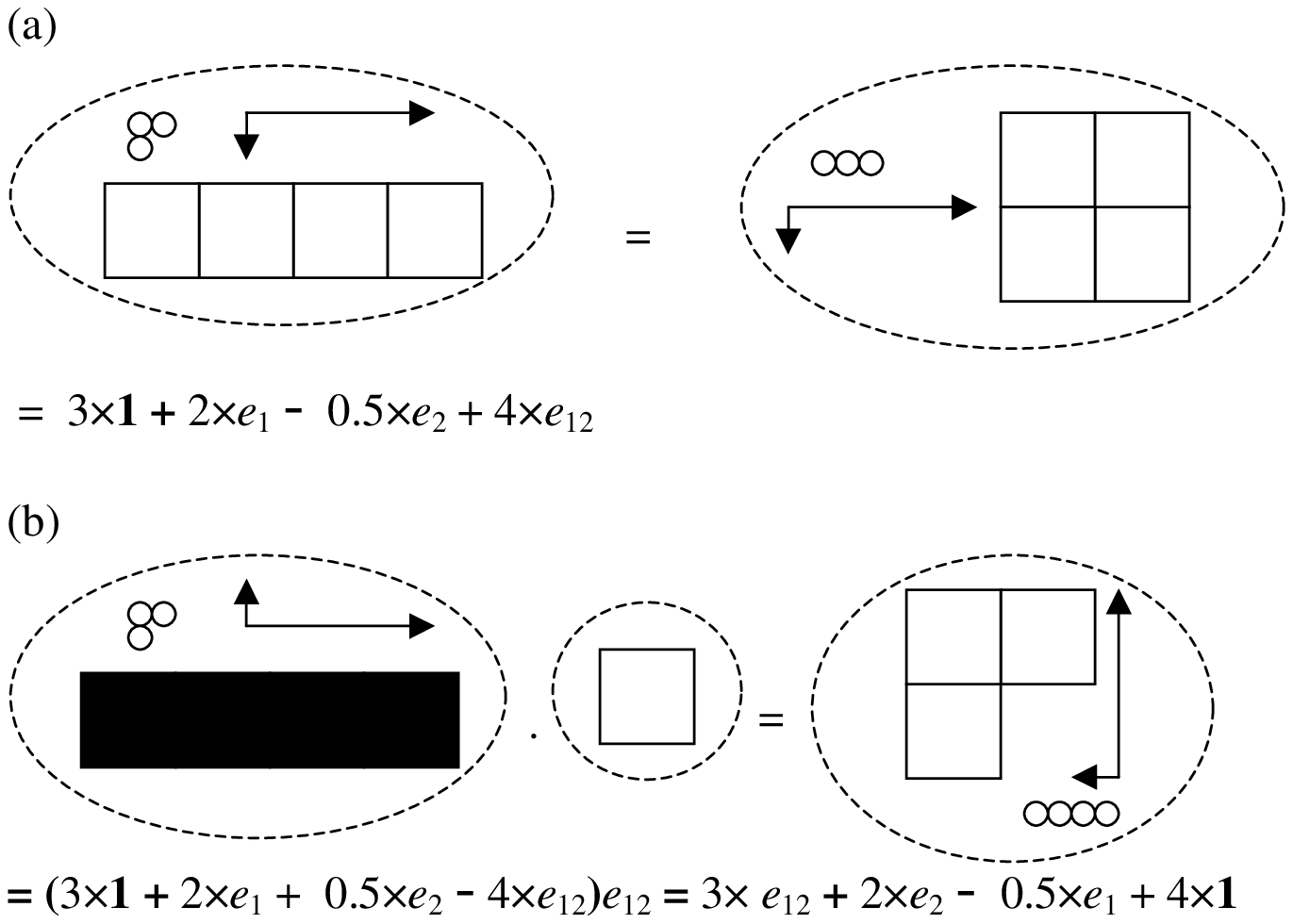}
\caption{(a) Multivectors are `bags of blades'. The two different
bags are equivalent. (b) Geometric product of a multivector and a
blade.}
\end{figure}
\begin{figure}
\includegraphics[width=8 cm]{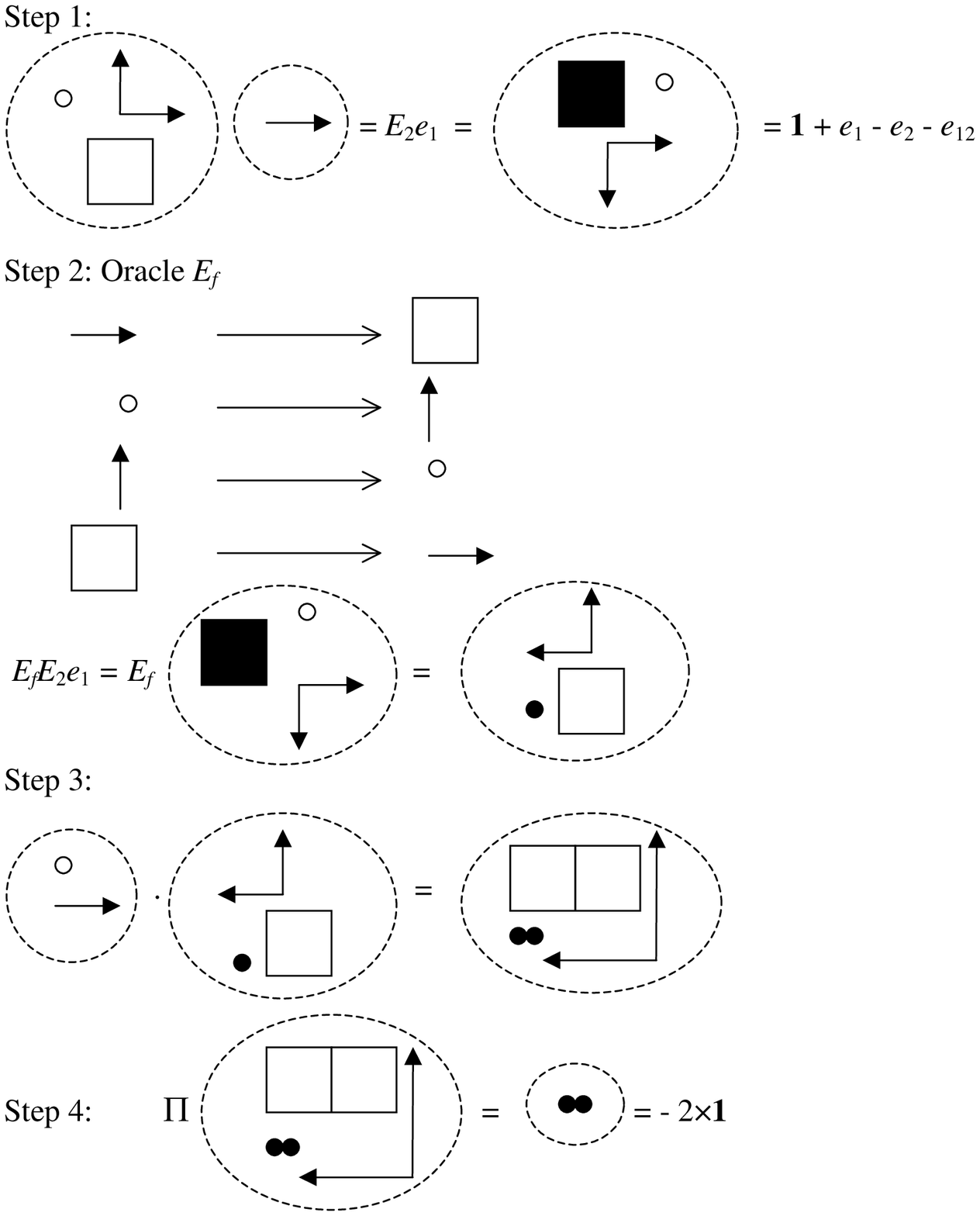}
\caption{DJ problem solved exclusively by means of
geometric operations: a 2-bit problem with $f(0)=f(1)=1$. One can {\it see\/} the result: The
two black dots mean $-2=(-1)^{f(0)}2^1$.}
\end{figure}
\begin{figure}
\includegraphics[width=6 cm]{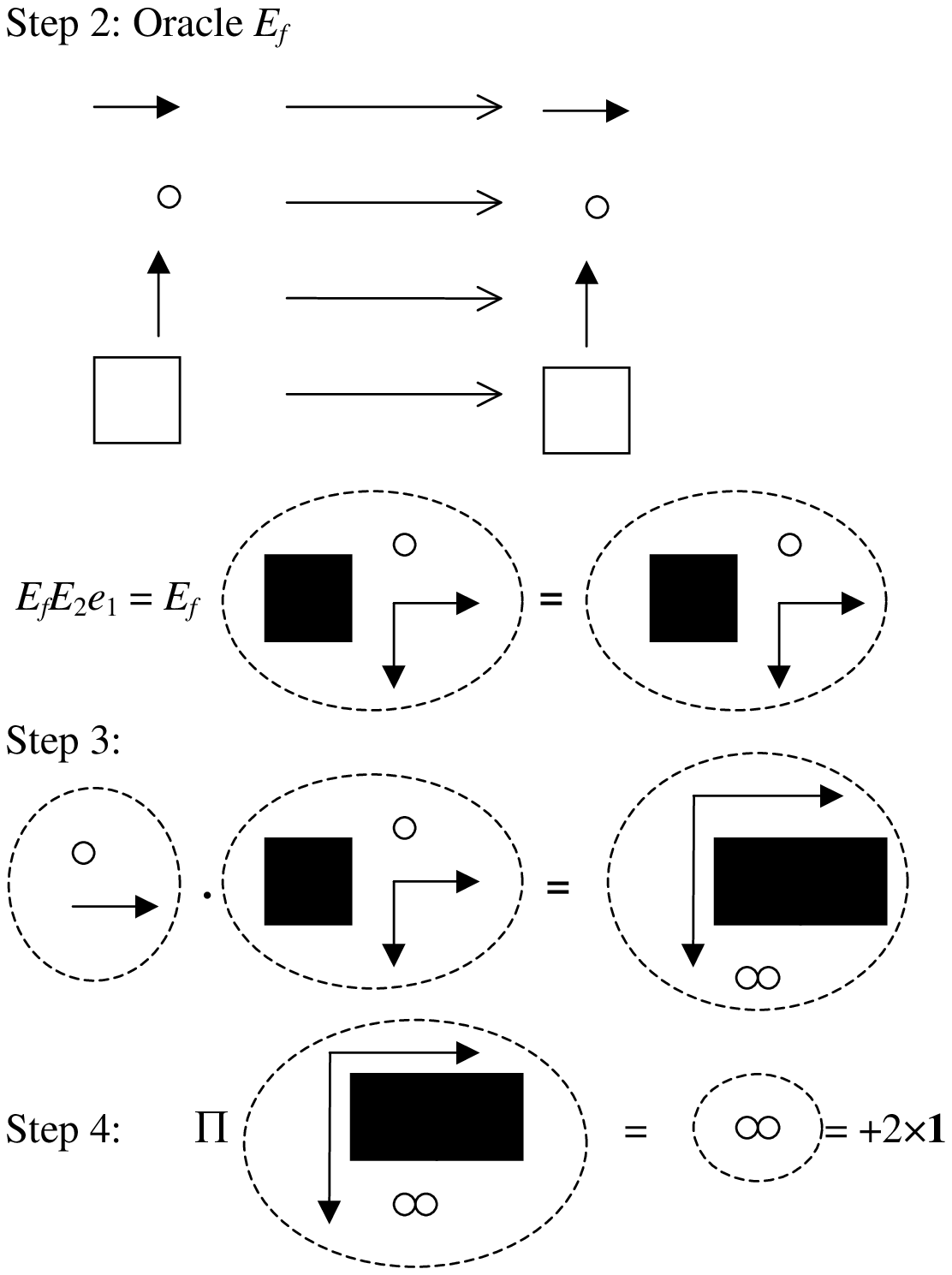}
\caption{The 2-bit problem and the constant function $f(0)=f(1)=0$.
The oracle acts trivially. One again sees the result: The two white
dots mean $+2=(-1)^{f(0)}2^1$.}
\end{figure}
\begin{figure}
\includegraphics[width=8 cm]{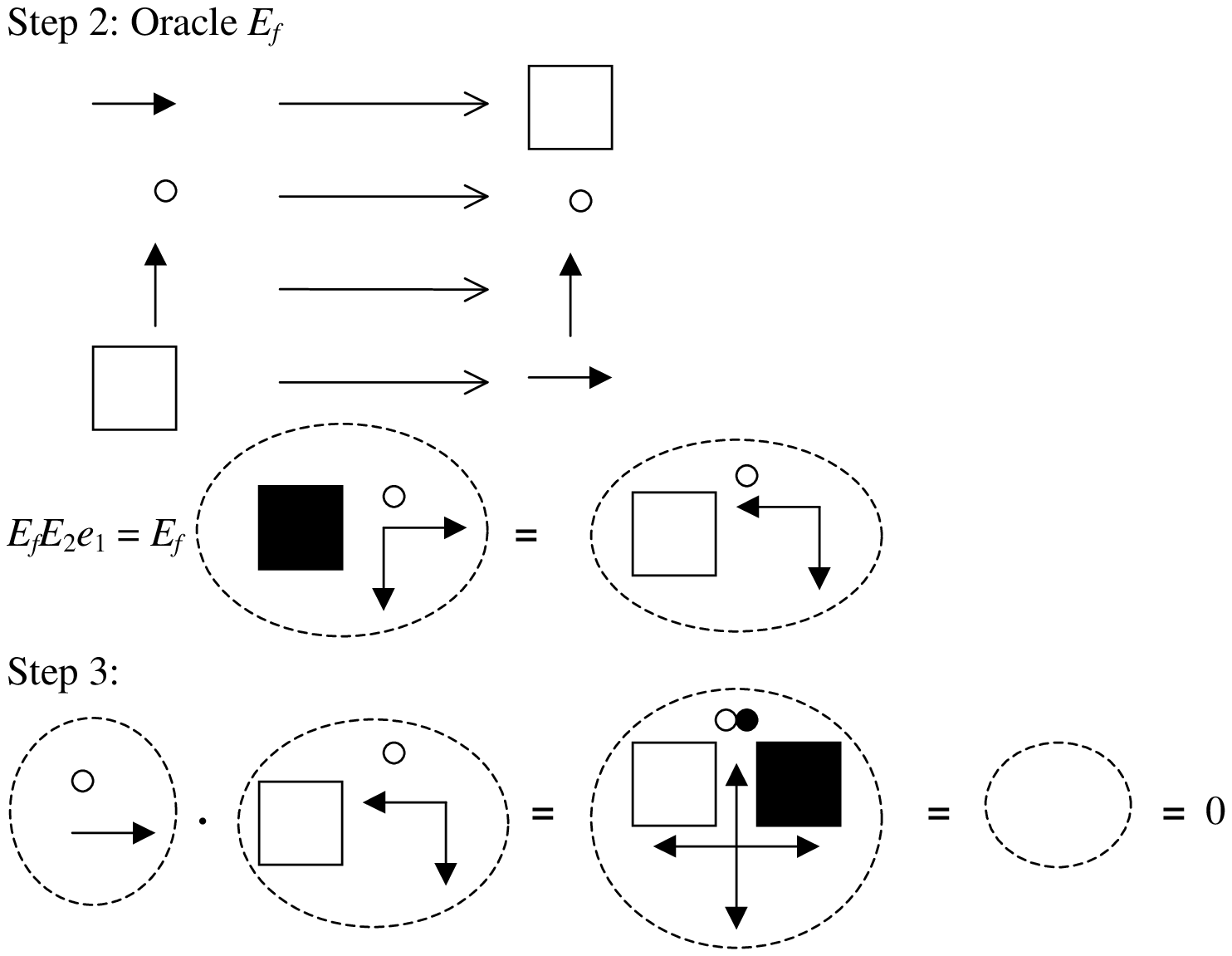}
\caption{The 2-bit problem and the balanced function $f(0)=0$,
$f(1)=1$. The bag is empty.}
\end{figure}
\begin{figure}
\includegraphics[width=6 cm]{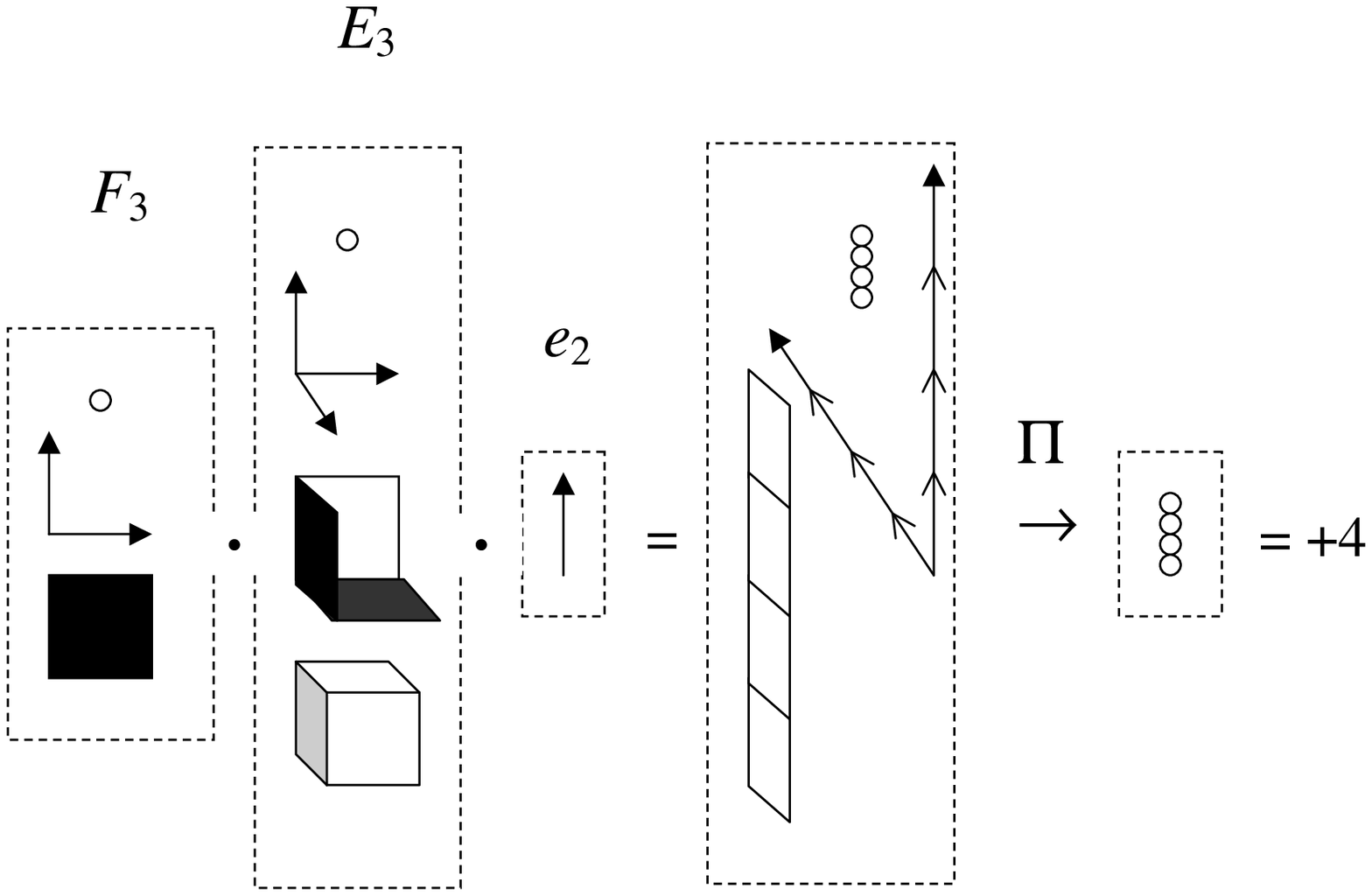}
\caption{3-bit problem for the constant function $f(00)=0$. The oracle is trivial. Four white dots mean $+4=(-1)^{f(00)}2^2$.}
\end{figure}


\begin{thebibliography}{99}
\bibitem{reverse}The reverse acts on blades as follows: $$
(e_{k_1}e_{k_2}\dots e_{k_{j-1}}e_{k_j})^{\dag}=
e_{k_j}e_{k_{j-1}}\dots e_{k_2}e_{k_1}, $$ and is extended to
multivectors by linearity.
\end{thebibliography}
\end{document}